\documentclass[journal=nalefd,manuscript=letter]{achemso}
\usepackage[T1]{fontenc}
\usepackage{amsmath,amssymb,bm}
\usepackage{graphicx}

\setkeys{acs}{articletitle=true,doi=true}
\usepackage{setspace}
\makeatletter
\renewcommand*{\acs@keywords@print}{%
  \ifx\@keywords\@empty\else
    \par\medskip
    \begingroup\small\setstretch{1}\raggedright%
      \noindent\textbf{Keywords: }\@keywords\par
    \endgroup
  \fi
}
\makeatother

\usepackage[hidelinks]{hyperref}
\newcommand{\unitsk}{\mathrm{W\,m^{-1}\,K^{-1}}}
\title{Heat Transport Mechanisms in MOF-5}
\author{Zezhu Zeng}
\email{zezhu.zeng@chem.ox.ac.uk}
\affiliation{Physical and Theoretical Chemistry Laboratory, Department of Chemistry, University of Oxford, South Parks Road, Oxford OX1 3QZ, United Kingdom}
\author{Xia Liang}
\affiliation{Department of Materials, Imperial College London, South Kensington Campus, London SW7 2AZ, United Kingdom}
\author{Hengtai Zheng}
\affiliation{Physical and Theoretical Chemistry Laboratory, Department of Chemistry, University of Oxford, South Parks Road, Oxford OX1 3QZ, United Kingdom}
\author{Aron Walsh}
\affiliation{Department of Materials, Imperial College London, South Kensington Campus, London SW7 2AZ, United Kingdom}
\author{David E. Manolopoulos}
\affiliation{Physical and Theoretical Chemistry Laboratory, Department of Chemistry, University of Oxford, South Parks Road, Oxford OX1 3QZ, United Kingdom}

\begin{document}
\begin{abstract}

We find that MOF-5 supports both long-lived phonons and linker vibrations whose split spectra signal a breakdown of the phonon quasiparticle approximation. 
Combining mode-projected molecular dynamics spectra with spectral Wigner heat transport, we resolve two contrasting transport channels: 
low-frequency particle-like phonon propagation decreases strongly from 200 to 500~K, whereas wave-like phonon coherence remains nearly constant and supplies about half the intrinsic conductivity at 500~K. 
A phenomenological extrinsic scattering model provides a plausible explanation for the discrepancy between calculated and measured conductivities in terms of both its magnitude and temperature dependence. 
We also identify signatures of phonon quasiparticle breakdown in the linker vibrations of ZIF-8 and HKUST-1, indicating that this feature extends to other metal--organic frameworks.

\end{abstract}
\clearpage
\noindent\textit{Introduction}.---
Our ability to assemble crystals from molecular building blocks makes crystal structure a chemical design variable~\cite{Hoskins1989Frameworks}.
Metal--organic frameworks (MOFs)~\cite{Yaghi1995Guests,Li1999,Eddaoudi2002Isoreticular} perfectly exemplify this principle: their metal nodes and organic linkers create pore environments that can be tailored for selective gas adsorption~\cite{Kondo1997Channels,Bloch2012Separations}.
Because adsorption--desorption cycles involve heat release and uptake, understanding microscopic heat transport in MOFs is important for optimizing their performance~\cite{Mason2015ThermalManagement}.

How heat is carried in a crystal depends strongly on how its atoms are bonded~\cite{Togo2015PhononLifetimes,Lindsay2013AbInitio,Broido2007Intrinsic,Zeng2025PNAS,Wieser2021Bottleneck}. 
MOFs complicate this picture because their organic linkers introduce molecular motions that may carry heat differently from collective framework vibrations~\cite{Nguyen2025MOFReview,Ryder2014THzMOF,Andreeva2020SoftMode,Yuan2026LinkerRotation,Islamov2023HighThroughput}.
Simulations have identified diverse heat carriers, from long-mean-free-path phonons in HKUST-1~\cite{Fan2022HKUSTPhonons} to nonpropagating vibrations in ZIF-4 and ZIF-62~\cite{Zhou2022ZIFTransport,Zhang2025ZIFThermalTransport}. Transport also largely depends on pore geometry~\cite{Babaei2017PoreGeometry} and adsorbed guests, which add phonon scattering~\cite{Babaei2020Adsorbates} and guest--host vibrational hybridization~\cite{DeCoster2022Hybridization}.
MOF-5, a cubic network of Zn$_4$O nodes and simple benzene-1,4-dicarboxylate (BDC) linkers, provides an archetypal system for examining how framework and linker motions govern heat transport. Yet a consistent microscopic picture of heat transport in MOF-5 remains elusive.

Huang {\em et al.} measured a room-temperature thermal conductivity ($\kappa$) of $0.32~\unitsk$ with an unusually weak, glass-like temperature dependence above 100~K~\cite{Huang2007}, and reproduced this behavior using Green--Kubo molecular dynamics (MD) simulations with an empirical potential~\cite{Huang2007MD}. 
By contrast, first-principles Boltzmann transport equation (BTE) calculations have since predicted a crystalline-like decrease of $\kappa$ with temperature~\cite{Zhang2021MOF5BTE}. 
More recent studies using machine-learned potentials have not established a consistent picture of either the magnitude or the temperature dependence of the thermal conductivity. 
Using a neuroevolution potential (NEP)~\cite{Fan2021}, Ying {\em et al.}\ found pronounced simulation-size effects and a converged $\kappa$ of $0.62~\unitsk$ at 300~K, nearly twice the experimental value, again decreasing with temperature~\cite{Ying2023}, whereas Wieser and Zojer obtained $0.32~\unitsk$ at 300~K from approach-to-equilibrium simulations with a moment-tensor potential~\cite{Wieser2024MLP}.
These discrepancies concern not only the magnitude and temperature dependence of $\kappa$, but also the heat transport mechanism~\cite{Lindner2026MOFHeatTransport}. Resolving this mechanism requires a mode-resolved description. 

In a lattice dynamics description of heat transport, the BTE describes heat conduction by intraband phonon population transport~\cite{Cahill2014Nanoscale,Minnich2015PhononTransport,He2012BulkThermalTransport}, while the Wigner transport equation (WTE)~\cite{Simoncelli2019Unified,Simoncelli2022Wigner,Legenstein2025OrganicTransport,Lindner2026MOFHeatTransport} and quasiharmonic Green--Kubo (QHGK) approaches~\cite{Isaeva2019QHGK,Barbalinardo2020QHGK,Fiorentino2023QHGK} also include  wave-like interband phonon coherence. 
For complex ultralow-$\kappa$ materials such as MOF-5, however, these approaches face two limitations. First, four-phonon~\cite{Feng2016FourPhonon,Feng2017FourPhonon,Guo2024SamplingPhonon} and higher-order~\cite{Xia2025FiveSixPhonon} scattering can become important but are computationally demanding for a system with a 106-atom primitive cell~\cite{Guo2024SamplingPhonon}. 
Second, describing each mode by a frequency and a linewidth becomes ill defined once the phonon spectra are non-Lorentzian~\cite{Paulatto2015Spectral,Monacelli2021SSCHA,Dangic2021Spectral,Fransson2023Overdamped,Zeng2026Spectral,Dangic2025Overdamped,Caldarelli2022,Castellano2025ModeCoupling}. 
MD overcomes both limitations: it includes anharmonicity to all orders~\cite{Reicht2026CrystallinePolymers}, and projecting MD trajectories onto phonon modes yields their full spectral line shapes~\cite{Sun2010Phonon,McGaughey2006PhononTransport}.

Here we combine MD-based phonon spectral densities~\cite{Zeng2026Spectral} with the spectral WTE~\cite{Caldarelli2022} to determine the heat transport mechanisms in MOF-5 beyond conventional BTE and standard WTE descriptions.
Retaining the full anharmonic spectra resolves population and coherence transport without imposing phonon quasiparticle approximations, revealing substantial coherence arising from spectral overlap already at 200~K.
We then incorporate a phenomenological extrinsic-scattering model into the spectral WTE to reconcile intrinsic calculations with measured conductivities.
Mode-resolved dynamics further reveal that long-lived phonons coexist with linker-dominated vibrations for which the quasiparticle picture breaks down.
Our results provide a comprehensive microscopic understanding of heat transport in MOF-5 and a framework for investigating other MOFs.

\noindent\textit{Theory}.---
We evaluate $\kappa$ using the spectral Wigner transport equation in the dressed-bubble approximation~\cite{Caldarelli2022}.
This is usually presented in terms of the phonon spectral density $b_{\bm qs}(\omega)$ for wave vector $\bm q$ and branch $s$, which one can write in terms of the corresponding Kubo spectrum $C^K_{\bm qs}(\omega)$ as~\cite{Zeng2026Spectral}
\begin{equation}
b_{\bm qs}(\omega) = \frac{\beta\hbar\omega}{2\pi}C^{K}_{\bm qs}(\omega)
\label{eq:bqs}
\end{equation}
where $\beta=1/k_{\rm B}T$. Here $C^K_{\bm qs}(\omega)$ is the Fourier transform $C^{K}_{\bm qs}(\omega)=\int_{-\infty}^{\infty} e^{i\omega t}C^{K}_{\bm qs}(t)\,{\rm d}t$ of the Kubo-transformed phonon amplitude correlation function
\begin{equation}
C^{K}_{\bm qs}(t) = \frac{1}{\beta}\int_0^{\beta} \langle \hat{a}_{\bm qs}(t-i\lambda\hbar)\hat{a}_{\bm qs}^{\dagger}(0)\rangle\,{\rm d}\lambda,
\label{eq:CKt}
\end{equation}
in which the angular brackets denote a quantum Boltzmann average.
Using Eq.~\eqref{eq:bqs}, the spectral WTE of Caldarelli {\em et al.}~\cite{Caldarelli2022} becomes
\begin{align}
\kappa^{ab} &= \frac{(\beta\hbar)^2}{4\pi N_{\bm q}V} \sum_{\bm qss'}\frac{(\omega_{\bm qs}+\omega_{\bm qs'})^2}{4}
v_{ss'}^a(\bm q)v_{s's}^b(\bm q)\nonumber\\
&\times \int_{-\infty}^{\infty} C^{K}_{\bm qs}(\omega)C^{K}_{\bm qs'}(\omega)c(\beta\hbar\omega)\,{\rm d}\omega,
\label{eq:kappa}
\end{align}
where $N_{\bm q}$ is the number of wave vectors, $V$ is the unit cell volume, $\omega_{\bm qs}$ is a harmonic phonon frequency, $v_{ss'}^{a}(\bm q)$ is a Wigner velocity matrix element, and $c(x)=k_{\rm B}(x/2)^2/[\sinh(x/2)]^2$ is the quantum harmonic heat capacity at $x=\beta\hbar\omega$. The diagonal ($s=s'$) and off-diagonal ($s\neq s'$) terms in Eq.~\eqref{eq:kappa} give the population ($\kappa_{\rm p}$) and coherence ($\kappa_{\rm c}$) contributions to the conductivity, with $\kappa_{\rm tot}=\kappa_{\rm p}+\kappa_{\rm c}$. In the phonon quasiparticle limit, the equation reduces to the lifetime-based WTE~\cite{Simoncelli2019Unified}.

We evaluate the Kubo spectrum $C^{K}_{\bm qs}(\omega)$ using the classical approximation to $C^{K}_{\bm qs}(t)$ introduced in ref~\citenum{Zeng2026Spectral}. The phonon annihilation operator is
$\hat a_{\bm qs}=(2\hbar\omega_{\bm qs})^{-1/2}(\omega_{\bm qs}\hat u_{\bm qs}+i\hat p_{\bm qs})$, where $\hat u_{\bm qs}$ and $\hat p_{\bm qs}=\dot{\hat u}_{\bm qs}$ are the mass-weighted displacement and momentum operators. Replacing these operators by the corresponding classical variables gives the mode-projected MD correlation function
\begin{equation}
C^{K}_{\bm qs}(t) \approx C^{\rm MD}_{\bm qs}(t)
=\langle a_{\bm qs}(t)a^*_{\bm qs}(0)\rangle,
\label{eq:CMD}
\end{equation}
where $a_{\bm qs}=(2\hbar\omega_{\bm qs})^{-1/2}
(\omega_{\bm qs}u_{\bm qs}+ip_{\bm qs})$.
This classical approximation is widely used in phonon normal-mode analyses~\cite{Sun2010Phonon,McGaughey2006PhononTransport,McGaughey2014Phonon}; it is exact in the harmonic limit, and it becomes exact in the high-temperature classical limit irrespective of the strength of anharmonicity~\cite{Zeng2026Spectral}. Since it is best justified for anharmonic solids above their Debye temperatures, we shall confine our attention to the 200--500~K range, which lies well above the \(\sim102\)~K Debye temperature of MOF-5~\cite{Huang2007MD,Huang2007}.

\begin{figure}[htbp]
\centering
\includegraphics[width=0.83\linewidth]{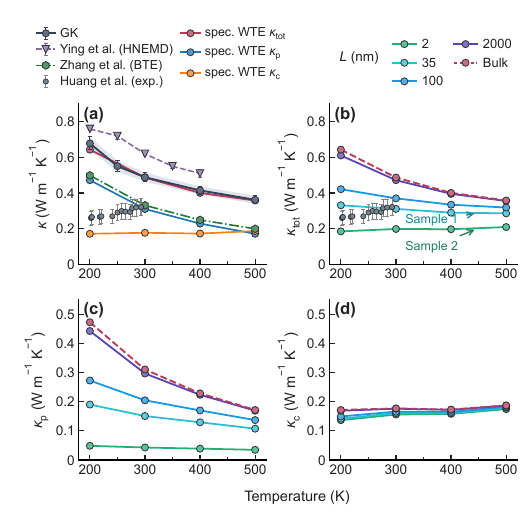}
\caption{Temperature-dependent thermal conductivity of MOF-5. (a) Green--Kubo and intrinsic spectral Wigner results compared with HNEMD~\cite{Ying2023}, first-principles BTE~\cite{Zhang2021MOF5BTE}, and measurements on four crystals~\cite{Huang2007}. 
(b--d) Total ($\kappa_{\rm tot}$), population ($\kappa_{\rm p}$), and coherence ($\kappa_{\rm c}$) spectral Wigner conductivities, respectively, for different effective extrinsic scattering lengths $L$. 
Panels (b--d) share the $L$ legend; dashed curves indicate the intrinsic bulk limit. Arrows in (b) indicate $L=35$ and 2 nm, inferred from the mean-free-path data for samples 1 and 2, respectively, of Huang {\em et al.}~\cite{Huang2007}. The former is a lower bound based on the unsaturated mean free path at 6 K, while the latter is its value in a low-temperature plateau.}
\label{fig:kappa}
\end{figure}

\noindent\textit{Thermal Conductivity}.---
We trained a NEP~\cite{Fan2021} on the high-quality MOF-5 DFT reference dataset of Wieser and Zojer~\cite{Wieser2024MLP}; computational details are given in the Supporting Information (SI). Figure~\ref{fig:kappa}a shows the isotropic conductivity, $\kappa=\mathrm{Tr}(\bm{\kappa})/3$, computed from Green--Kubo and spectral WTE calculations with our NEP. The two agree closely from 300 to 500~K. At 200~K, the Green--Kubo calculation gives a slightly higher $\kappa$, consistent with its use of the classical heat capacity rather than the quantum one in the integrand of Eq.~\eqref{eq:kappa}. Both calculations show a crystalline-like decrease of $\kappa$ between 200 and 500~K, consistent with homogeneous nonequilibrium MD (HNEMD) results of Ying {\em et al.} obtained with a NEP trained on a different dataset~\cite{Ying2023}.

The spectral WTE decomposition shows that the decrease in $\kappa$ arises almost entirely from the population term: $\kappa_{\rm P}$ falls from 0.47 to 0.17~$\unitsk$ between 200 and 500~K, whereas $\kappa_{\rm C}$ remains nearly constant at 0.17--0.19~$\unitsk$. Consequently, $\kappa_{\rm tot}$ decreases from 0.64~$\unitsk$ at 200~K to 0.36~$\unitsk$ at 500~K, while the coherence fraction rises from 26.6\% to 52.2\%, slightly exceeding the population contribution at the highest temperature. The decrease in $\kappa_{\rm P}$ is consistent with the crystalline-like temperature dependence predicted by the earlier first-principles BTE calculation~\cite{Zhang2021MOF5BTE}, which includes only the population contribution.

Both our Green--Kubo and spectral WTE calculations overestimate the measured thermal conductivity of Huang {\em et al.}~\cite{Huang2007} (Fig.~\ref{fig:kappa}a), as do the earlier HNEMD calculations of Ying et al.~\cite{Ying2023}. However, these calculations describe intrinsic heat transport, whereas the experimental conductivity is also affected by extrinsic scattering processes. To assess whether these processes can account for the discrepancy between theory and experiment, we shall now  incorporate extrinsic scattering into the spectral WTE using an experimentally motivated phenomenological model.

Huang {\em et al.}\ measured the thermal conductivities of four MOF-5 crystals prepared under different solvent conditions. Their low-temperature analysis compared the crystals with the highest and lowest conductivities, which they labeled samples~1 and 2, respectively~\cite{Huang2007}. For each crystal, they estimated an average mean free path, $\ell_{\rm eff}(T)=3\kappa(T)/[c_V(T)\bar v]$, using the Debye heat capacity per unit volume $c_V(T)$ and average phonon velocity $\bar v$. We can interpret this effective mean free path using Matthiessen's rule for independent resistive processes~\cite{Callaway1959Thermal}, $\ell_{\rm eff}^{-1}(T)\simeq\ell_{\rm int}^{-1}(T)+L^{-1}$, where $\ell_{\rm int}$ describes intrinsic phonon--phonon scattering and $L$ is a temperature-independent effective extrinsic length. At sufficiently low temperature, where $\ell_{\rm int}\gg L$, the effective mean free path approaches $L$. Sample~2 exhibited a plateau near 2~nm, attributed by Huang {\em et al.}\ to defects, whereas sample~1 reached approximately 35~nm at 6~K without saturating~\cite{Huang2007}. Thus, 2~nm provides an estimate of $L$ for sample~2, whereas 35~nm provides only a lower bound on $L$ for sample~1.

We can incorporate this extrinsic scattering in a phenomenological model. For each mode, the extrinsic energy relaxation time is $\tau^L_{\bm qs}=L/|\bm v_{ss}(\bm q)|$, where $\bm v_{ss}(\bm q)$ is the diagonal group velocity vector.
Assuming that extrinsic scattering provides an independent exponential energy-relaxation channel for each mode, the corresponding phonon amplitude decays at half the energy relaxation rate, giving
$C^{L}_{\bm qs}(t)=C_{\bm qs}^{\rm MD}(t)e^{-|t|/2\tau^L_{\bm qs}}$.
By the Fourier convolution theorem, this additional damping convolves the intrinsic correlation spectrum with a Lorentzian to give
\begin{equation}
    C^L_{\bm qs}(\omega) = \frac{\eta^L_{\bm qs}}{\pi} \int_{-\infty}^{\infty} \frac{C^{\rm MD}_{\bm qs}(\omega')}{(\omega-\omega')^2+(\eta^L_{\bm qs})^2}\,{\rm d}\omega',
\end{equation}
where $\eta^L_{\bm qs}=1/2\tau^L_{\bm qs}=|\bm v_{ss}(\bm q)|/2L$. The modified spectrum $C^L_{\bm qs}(\omega)$ can again be used in place of $C^K_{\bm qs}(\omega)$ in Eq.~\eqref{eq:kappa}. Note that there is no analogous (mode-resolved) way to incorporate extrinsic scattering in the Green-Kubo formulation.

Including the extrinsic scattering in Eq.~\eqref{eq:kappa} gives conductivities that bracket the measurements from all four of Huang {\em et al.}'s samples between 200 and 300~K (Fig.~\ref{fig:kappa}b). At 300~K, $L=35$~nm gives $0.31~\unitsk$, close to the measured $0.32~\unitsk$, whereas $L=2$~nm gives $0.20~\unitsk$. These two values of $L$ were inferred from the low-temperature data for samples 1 and 2; our calculations show that they bracket all four experimental measurements at higher temperatures.
Extrinsic scattering mainly suppresses population transport: at 300~K and $L=35$~nm, $\kappa_p$ falls from 0.31 to 0.15~$\unitsk$, whereas $\kappa_c$ changes only from 0.18 to 0.16~$\unitsk$ (Fig.~\ref{fig:kappa}c,d).
This selective suppression provides an explanation for both the lower conductivity and its weak temperature dependence in the experiment.
At $L=2000$~nm, $\kappa$ approaches the intrinsic limit, suggesting that samples with micrometer-scale effective scattering lengths would be needed to reveal the intrinsic conductivity.

\begin{figure}[htbp]
\centering
\includegraphics[width=\linewidth]{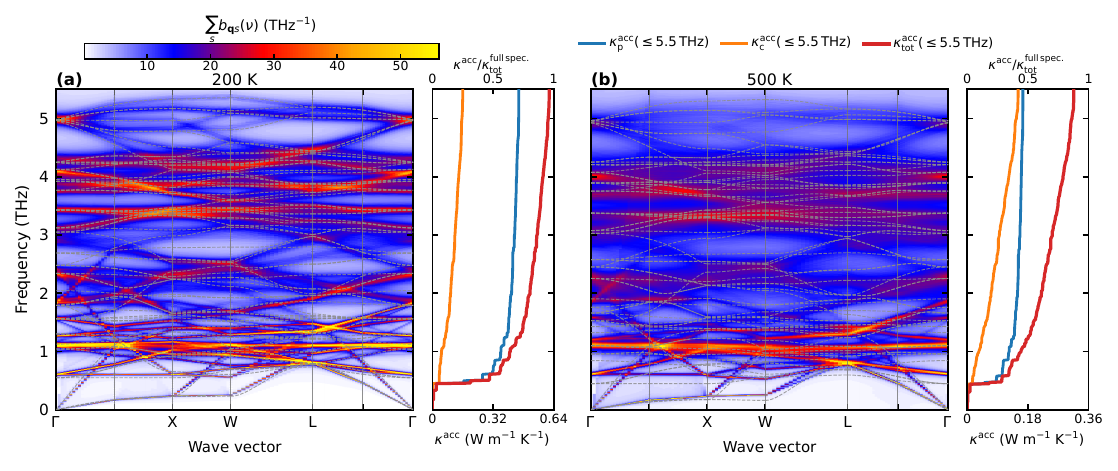}
\caption{Frequency-resolved heat transport in MOF-5 at (a) 200 and (b) 500~K.
The heatmaps show the summed mode spectral density along $\Gamma$--X--W--L--$\Gamma$, interpolated between sampled wave vectors; dashed curves are harmonic branches evaluated at the cell volume for each temperature. The adjacent curves give the accumulated population (blue), coherence (orange), and total (red) conductivities as a function of harmonic frequency up to 5.5~THz. The upper horizontal axes on these panels give these accumulated conductivities divided by the total conductivity of the full spectrum.}
\label{fig:spectral}
\end{figure}

\noindent\textit{Phonon Spectra}.---
Figure~\ref{fig:spectral} shows the summed MD spectral density along a high-symmetry path through the Brillouin zone together with the harmonic phonon dispersions. For visualization, the spectral density is interpolated from the $4\times4\times4$ $\bm q$-mesh used in the transport calculations (see SI for interpolation details). The harmonic dispersions are evaluated at the temperature-dependent cell volumes, thereby accounting for the negative thermal expansion of MOF-5~\cite{Lock2010MOF5NTE,Rimmer2014AcousticNTE,Zhou2008MOF5NTE}.

Heating from 200 to 500~K produces a nonuniform spectral response: the closely spaced 3--4.5~THz ridges broaden and lose contrast, whereas low-frequency features, including the weakly dispersive branch between X and W, remain resolved (Fig.~\ref{fig:spectral}a,b). The lowest-frequency features near X, W, and L lie above their harmonic frequencies, especially at 500~K, while several higher-frequency features soften on heating. Since the harmonic branches are evaluated at the corresponding equilibrium cell volumes, these upward shifts reflect anharmonic hardening beyond the quasiharmonic effect of thermal contraction.

The accumulation curves in Fig.~\ref{fig:spectral} reveal distinct frequency ranges for the two transport channels. Population conductivity accumulates mainly between 0.3 and 1.2~THz and then nearly saturates, whereas coherence conductivity continues to grow through the dense higher-frequency bands. At 200 and 500~K, modes with harmonic frequencies below 1~THz supply 77\% and 72\% of $\kappa_{\rm p}$, respectively, but only 29\% and 19\% of $\kappa_{\rm c}$. Population transport is therefore concentrated in a narrow low-frequency window, whereas coherent transport extends over a broader frequency range. Despite this broader range, modes below 5.5~THz account for 96.3\% and 87.9\% of the total conductivity at 200 and 500~K, respectively; transport above this window is predominantly coherent and becomes relatively more important on heating.

\begin{figure}[htbp]
\centering
\includegraphics[width=\linewidth]{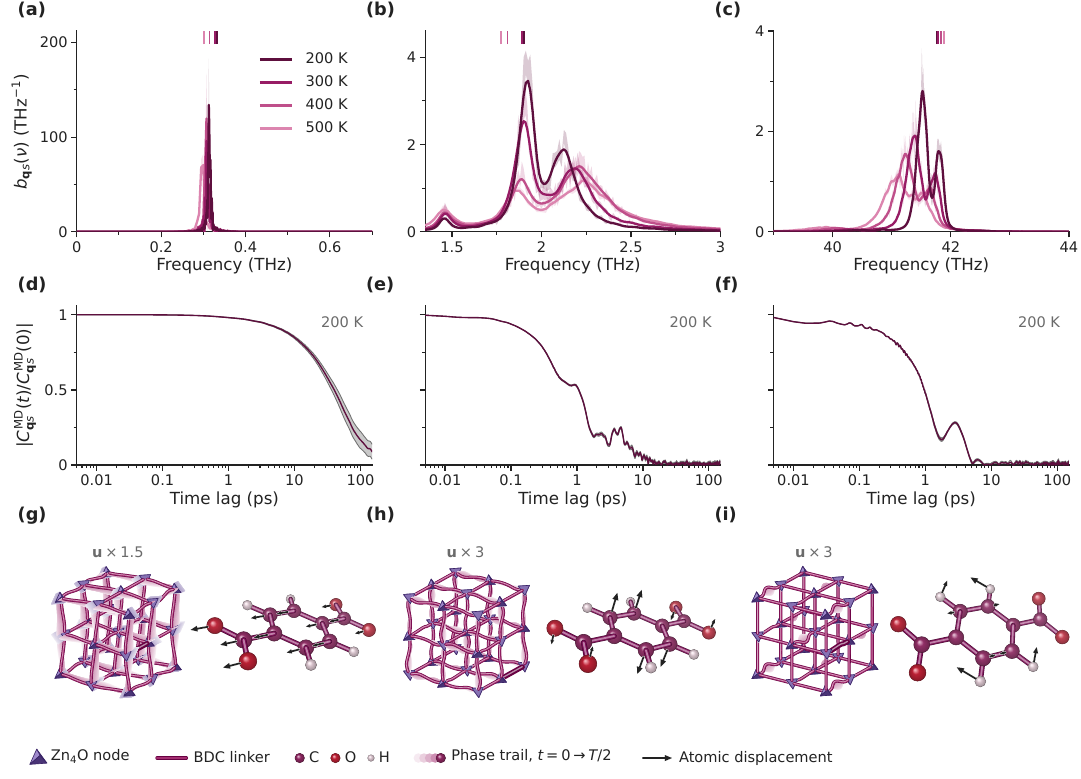}
\caption{Contrasting vibrational dynamics in MOF-5. Columns correspond to the collective acoustic mode $\bm q=(0.50,0.25,0.25)$, $s=0$; linker libration $\bm q=(0.75,0.50,0.75)$, $s=22$; and in-plane linker deformation $\bm q=(0.50,0.25,0.75)$, $s=257$.
(a--c) Spectral densities at 200--500~K. Short vertical marks indicate the harmonic frequencies at each temperature.
(d--f) Magnitude of the normalized correlation, $|C^{\rm MD}_{\bm qs}(t)/C^{\rm MD}_{\bm qs}(0)|$, at 200~K as a function of time lag on a logarithmic scale. Curves in (a--f) are means over ten MD runs per temperature, and shading denotes $\pm$ one standard error of the mean.
(g--i) Displacement patterns from harmonic eigenvectors. Tetrahedra represent Zn$_4$O nodes and ribbons represent BDC linkers. Fading copies trace half a vibrational period, with displacements magnified as indicated. Arrows on the linker atoms show nearly rigid translation in (g), benzene-ring libration in (h), and in-plane deformation in (i).}
\label{fig:modes}
\end{figure}

\noindent\textit{Vibrational Dynamics}.---
To relate the contrasting transport windows in Fig.~\ref{fig:spectral} to atomic motion, we compare a collective acoustic mode near 0.3~THz with a BDC-dominated optical mode near 2~THz, both within the main heat-carrying range (below 5.5~THz). A third mode near 42~THz involves motion concentrated on the linkers outside this range. Their participation ratios at 200~K, 0.985, 0.365, and 0.125, respectively, quantify the increasing spatial concentration of the displacements. Figure~\ref{fig:modes}g--i shows their phase-dependent displacement patterns, constructed from harmonic eigenvectors computed with Phonopy~\cite{Togo2023PhonopyImplementation}. The displacements are normalized independently for each mode and magnified by factors of 1.5, 3, and 3 for visibility; construction details and linker animations (Movies~S1-3) are provided in the SI.

In the acoustic mode, Zn$_4$O nodes and BDC linkers move together, collectively deforming the network (Fig.~\ref{fig:modes}g). The detailed linker view shows nearly parallel atomic displacements, indicating predominantly rigid translation. Its spectrum retains a single Lorentzian peak near 0.3~THz that softens and broadens on heating (Fig.~\ref{fig:modes}a). At 200~K, the correlation decays smoothly and remains appreciable beyond 100~ps (Fig.~\ref{fig:modes}d), illustrating the long-lived collective dynamics that underlie low-frequency population transport.

The linker-dominated modes involve distinct molecular motions but both have split spectra. Near 2~THz, the benzene ring undergoes torsional libration relative to the carboxylate groups, while the Zn$_4$O nodes move much less (Fig.~\ref{fig:modes}h). Peaks near 1.9 and 2.1~THz, accompanied by a weak feature near 1.45~THz, evolve into two lobes of comparable intensity on heating (Fig.~\ref{fig:modes}b). At 200~K, the correlation decays within a few picoseconds but shows shoulders and partial revivals (Fig.~\ref{fig:modes}e). Near 42~THz, the linker instead deforms predominantly within the benzene-ring plane, with nearly stationary nodes (Fig.~\ref{fig:modes}i). Peaks near 41.5 and 41.8~THz broaden and shift spectral weight downward on heating (Fig.~\ref{fig:modes}c), while the correlation briefly revives before nearly vanishing within 6~ps (Fig.~\ref{fig:modes}f). These split spectra signal a breakdown of the phonon quasiparticle picture, and the partial correlation revivals are consistent with beating between spectral components. Importantly, this behavior occurs not only in the high-frequency linker deformation but also in the low-frequency linker libration, which lies within the frequency range over which $\kappa_{\rm c}$ accumulates.

We have extended this analysis beyond MOF-5 to assess the phonon quasiparticle character of selected linker-dominated modes in ZIF-8 and HKUST-1 at 300~K (Fig.~S6 in the SI). The ZIF-8 modes exhibit strongly split spectra, whereas the HKUST-1 modes retain a sharp dominant resonance with broader secondary spectral weight. These results indicate that departures from the phonon quasiparticle picture are not unique to MOF-5 and reinforce the need to retain full spectral line shapes when evaluating heat transport in MOFs.

\noindent\textit{Conclusion}.---
In summary, mode-projected MD spectra combined with the spectral WTE provide a framework for calculating heat transport in MOFs without assuming phonon quasiparticles. In MOF-5, long-lived phonons coexist with linker vibrations that have split spectra; population transport decreases strongly between 200 and 500~K, whereas coherent transport remains nearly constant and supplies about half of the intrinsic $\kappa$ at 500~K. Phenomenological extrinsic scattering provides an explanation for the lower and more weakly temperature-dependent conductivity observed experimentally, mainly by suppressing population transport.\cite{Cai2026WaterDiffusion} Measurements on crystals with reduced extrinsic scattering could therefore reveal the higher intrinsic conductivity of MOF-5, and signatures of similar quasiparticle breakdown in the linker vibrations of ZIF-8 and HKUST-1 suggest that full spectral line shapes could be important for understanding heat transport in other MOFs.

\begin{suppinfo}
NEP training and validation, thermal expansion and harmonic phonons, Green--Kubo sampling and size dependence, spectral Wigner transport and correlation functions, spectral-map interpolation, mode visualization, and linker spectra in ZIF-8 and HKUST-1 (PDF); collective acoustic motion (Movie S1, GIF), BDC linker libration (Movie S2, GIF), and in-plane BDC linker deformation (Movie S3, GIF).
\end{suppinfo}

\bigskip
\noindent\textbf{Notes}\par
The authors declare no competing financial interest.

\begin{acknowledgement}
Z.Z. acknowledges support from a Royal Society Newton International Fellowship (Grant No. NIF/R1/254124). H.Z. acknowledges support from NCCR Separations, a National Centre of Competence in Research funded by the Swiss National Science Foundation (Grant No. 229280). The authors acknowledge the use of the University of Oxford Advanced Research Computing (ARC) facility.
\end{acknowledgement}

\bibliography{references}

% Insert the author-selected final TOC graphic on the last page.
\clearpage\noindent\textbf{For Table of Contents Only}\par\medskip
\noindent\includegraphics[width=3.25in,height=1.75in,keepaspectratio]{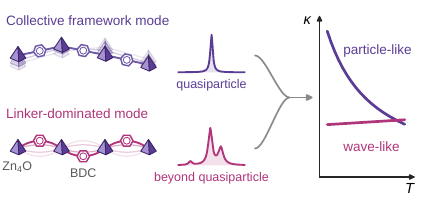}
\end{document}